\documentclass{article}

\usepackage{amssymb}

\usepackage{epsfig}

\usepackage{graphicx,psfrag,amsmath}

\begin{document}

\def\beq#1\eeq{\begin{equation}#1\end{equation}}
\def\beql#1#2\eeql{\begin{equation}\label{#1}#2\end{equation}}

\def\bea#1\eea{\begin{eqnarray}#1\end{eqnarray}}
\def\beal#1#2\eeal{\begin{eqnarray}\label{#1}#2\end{eqnarray}}

\newcommand{\Z}{{\mathbb Z}}
\newcommand{\N}{{\mathbb N}}
\newcommand{\C}{{\mathbb C}}
\newcommand{\Cs}{{\mathbb C}^{*}}
\newcommand{\R}{{\mathbb R}}
\newcommand{\intT}{\int_{[-\pi,\pi]^2}dt_1dt_2}
\newcommand{\cC}{{\mathcal C}}
\newcommand{\cI}{{\mathcal I}}
\newcommand{\cN}{{\mathcal N}}
\newcommand{\cE}{{\mathcal E}}
\newcommand{\cA}{{\mathcal A}}
\newcommand{\xdT}{\dot{{\bf x}}^T}
\newcommand{\bDe}{{\bf \Delta}}

\def\ket#1{\left| #1\right\rangle }
\def\bra#1{\left\langle #1\right| }
\def\braket#1#2{\left\langle #1\vphantom{#2}
  \right. \kern-2.5pt\left| #2\vphantom{#1}\right\rangle }
\newcommand{\gme}[3]{\bra{#1}#3\ket{#2}}
\newcommand{\ome}[2]{\gme{#1}{#2}{\mathcal{O}}}
\newcommand{\spr}[2]{\braket{#1}{#2}}
\newcommand{\eq}[1]{Eq\,\ref{#1}}
\newcommand{\xp}[1]{e^{#1}}

\def\limfunc#1{\mathop{\rm #1}}
\def\Tr{\limfunc{Tr}}

\def\dr{detector }
\def\drs{detectors }
\def\drn{detector}
\def\dtn{detection }
\def\dtnn{detection}

\def\pho{photon }
\def\phon{photon}
\def\phos{photons }
\def\phosn{photons}
\def\mmt{measurement }
\def\an{amplitude}
\def\a{amplitude }
\def\co{coherence }
\def\con{coherence}

\def\st{state }
\def\stn{state}
\def\sts{states }
\def\stsn{states}

\def\cow{"collapse of the wavefunction"}
\def\de{decoherence }
\def\den{decoherence}
\def\dm{dark matter }
\def\dmn{dark matter}

\newcommand{\mop}{\cal O }
\newcommand{\dt}{{d\over dt}}
\def\qm{quantum mechanics }
\def\qms{quantum mechanics }
\def\qml{quantum mechanical }

\def\qmn{quantum mechanics}
\def\mmtn{measurement}
\def\pow{preparation of the wavefunction }

\def\me{ L.~Stodolsky }
\def\T{temperature }
\def\Tn{temperature}
\def\t{time }
\def\tn{time}
\def\wfs{wavefunctions }
\def\wf{wavefunction }
\def\wfn{wavefunction} 
\def\wfsn{wavefunctions}
\def\wvp{wavepacket }
\def\pa{probability amplitude } 
\def\sy{system } 
\def\sys{systems }
\def\syn{system} 
\def\sysn{systems} 
\def\ha{hamiltonian }
\def\han{hamiltonian}
\def\rh{$\rho$ }
\def\rhn{$\rho$}
\def\op{$\cal O$ }
\def\opn{$\cal O$}
\def\yy{energy }
\def\yyn{energy}
\def\yys{energies }
\def\yysn{energies}
\def\pz{$\bf P$ }
\def\pzn{$\bf P$}
\def\pl{particle }
\def\pls{particles }
\def\pln{particle}
\def\plsn{particles}

\def\plz{polarization  }
\def\plzs{polarizations }
\def\plzn{polarization}
\def\plzsn{polarizations}

\def\sctg{scattering }
\def\sctgn{scattering}

\def\prob{probability }
\def\probn{probability}

\def\om{\omega} 

\def\hf{\tfrac{1}{2}}

\def\zz{neutrino }
\def\zzn{neutrino}
\def\zzs{neutrinos }
\def\zzsn{neutrinos}

\def\zn{neutron}
\def\zns{neutrons}

\def\csss{cross section }
\def\csssn{cross section}

\title{Features of Fast Neutrons in Dark Matter
Searches}

\author{ 
L. Stodolsky, \\
Max-Planck-Institut f\"ur Physik
(Werner-Heisenberg-Institut)\\
F\"ohringer Ring 6, 80805 M\"unchen, Germany}

\maketitle

\begin{abstract}
 Diffractive \sctg of ``fast'' or  ``high \yy '' \zns\, can give
low \yy nuclear recoils in
the signal region for \dm searches. We present a discussion
using the `black disc' model. This permits
a simple and general, although approximate, description of
this possible  background. We note a number of its features. In
particular there are mass number A dependent aspects which can
be studied in setups where events on different nuclei are
observable at the same time. These
include the recoil \yy distributions, and the  A behavior of the
\csssn. We define a parameter 
$E^o_R$ which characterizes the recoil \yy to be expected due to
fast \zns. It ranges from 100 keV on light nuclei to a few keV on
heavy nuclei, and a general treatment is possible in terms of it,
within  the `black disc' approximation.
In addition the presence of inelastic processes on the nuclei at
about the same level of
elastic processes would be characteristic of fast \zns.
\end{abstract}

\section{Introduction}
In the search
for \dm by direct detection in the laboratory, spectacular levels
of background reduction have been achieved. In the ``Commissioning
Run'' of the CRESST II \drn, for example, the rate of posssibly
interesting events was reported as 0.063 per kg-day \cite{cr2},
based on three events. For
the  about 600 gm of sensitive material used in the \dm analysis,
this corresponds to less than one event per week. Similarly, the
CDMS collaboration, after various cuts, had no events  for 121 kg-
days of germanium data \cite{cdms}. EDELWEISS discusses four events
on the basis of 322 kg-days \cite{edelw}. And most recently
XENON100, with 48 kg for 100 days, reports only three events
\cite{xenon}.

With such  low event rates even rare sources of background must be
considered. In particular 
the most  worrisome background in underground experiments searching
for \dm through nuclear recoils are  neutron-induced nuclear
recoils. Neutrons
scatter off nuclei and the recoils could 
 closely resemble the sought-for signal. Furthermore as we shall
explain below,  fast neutrons, which are difficult to shield, will
give elastic nuclear recoils with the same general \yys  as
expected in the \dm searches. Although estimates   of \zns\,
from standard sources such as incoming muons or radioactivity in
the surrounding rock ( see the above references and for example 
\cite{wul}) give \zn~fluxes which are quite small,  the very good
level of
background suppression being reached implies that
even very rare processes or perhaps unexpected contaminations must
be considered.
While there are elaborate Monte-Carlo program sets for carrying out
standard background calculations, it must be said that they are 
not very transparent,
and  do not offer much insight into the nature of the effects
under consideration. Hence it may be useful to have a discussion 
based on a simple physical picture, even if this
is only semi-quantitatively accurate.

 Estimates  generally lead to
neutron sprectra  dominated by  low \yysn.   
 Such fluxes can be strongly reduced by \zn\, shields. 
However such shields
are typically hydrocarbon materials, which while very effective at
low \yyn, become rapidly ineffective at higher \yysn. Thus the PDG
tables \cite{pdg} give the rather large value of  88\,cm for the
nuclear interaction length
in polyethylene, and for example in a Monte Carlo study with a
lead-hydrocarbon
shield \cite{ara} the authors  remark that this ``provides
effective shielding only for \zns\, below 10 MeV''. 

Thus there is a  possibly significant question of  ``fast''  or
``higher \yy''
\zns.
 It may therefore be useful if we present
some simple remarks about the effects to be anticipated from 
them in \dm \drn s and perhaps other low background setups. We can
do this  in a simple and general way
terms using the
`black disc' model which,  for fast \zns, approximately applies for
all nuclei.  While this can
only be regarded as a semi-quantative approximation, it offers a
simple and
general overwiew of the situation. Unfortunately,  even  such an 
rough discussion is not possible for slow \zns.

We should stress that we use the terms ``fast'' or ``high \yy''  in
the sense of \zn\, physics, that is to mean \zn s of a few MeV and
above, and of course not  in the sense of high \yy \pl physics.

\section{``High \yy'' neutrons and the ``Black disc''}

Above a few MeV \zn -nucleus \sctg
begins to resemble the ``black disc" limit. While at lower \yys the
behavior of the \csss is dominated by resonances and shows a
complicated structure  which varies from nucleus to nucleus, at the
higher \yys a smooth behavior with respect to 
\zn\, \yy sets in, one which can approximately
be described by the `black disc'model. This model in its simplest
form
 only involves one parameter,
namely R, the radius of the nucleus. This permits a simple
description, applicable to all target nuclei, adequate for our
purposes of semi-quantative background estimates.

\begin{figure}
\includegraphics[width=\linewidth]{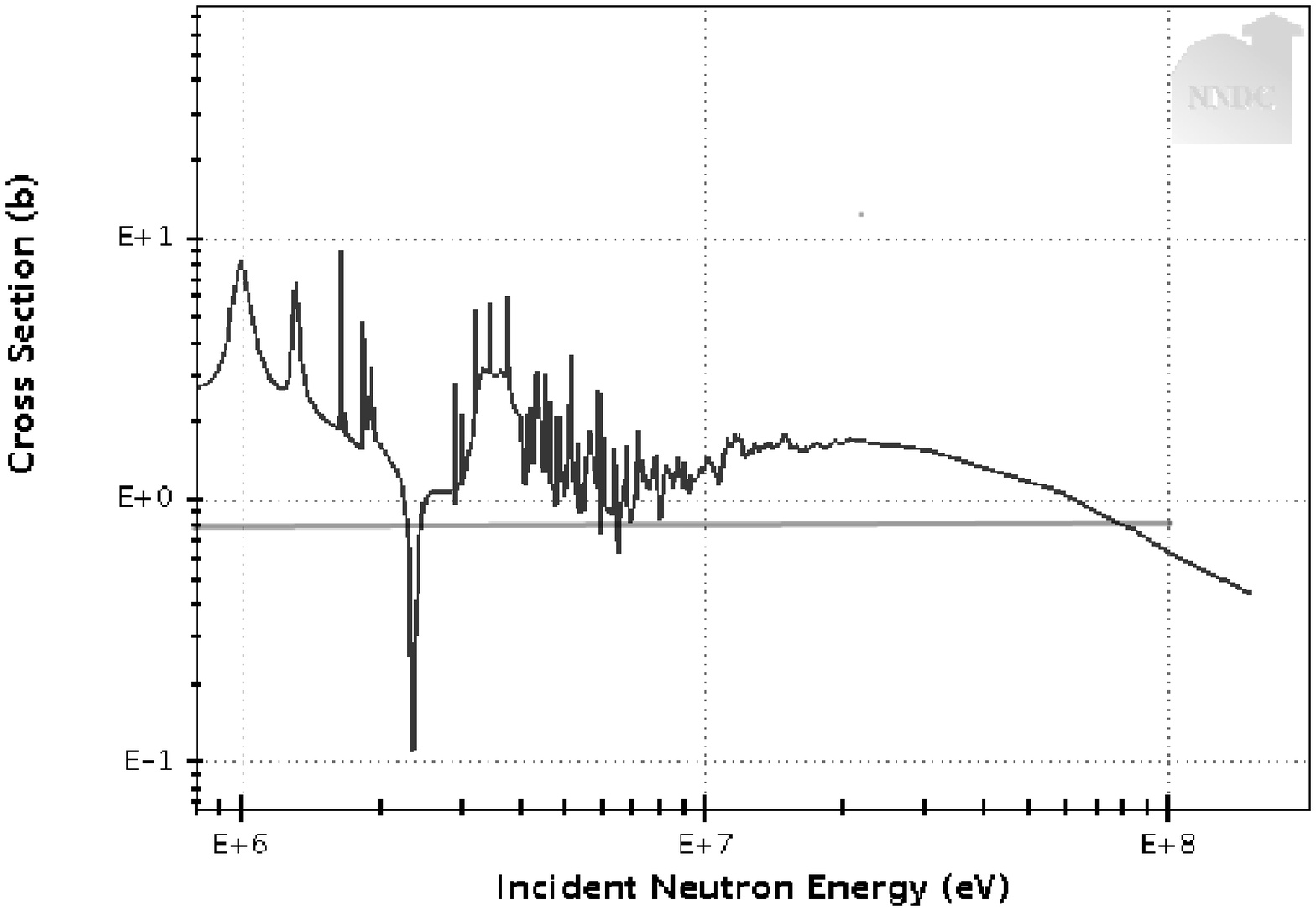}
\caption{ The total \csss for \zn s on $^{16}O$, from the data
compilations of the of the neutron data center NNDC at Brookhaven
National Laboratory.  Above the
resonance region the \csss behaves smoothly. The `black disc'
estimate (grey line), using $R=A^{1/3}1.4\,f$,  is $2\times \pi
R^2=0.78~b$.}
\label{o16tot}
\end{figure}

\begin{figure}
\includegraphics[width=\linewidth]{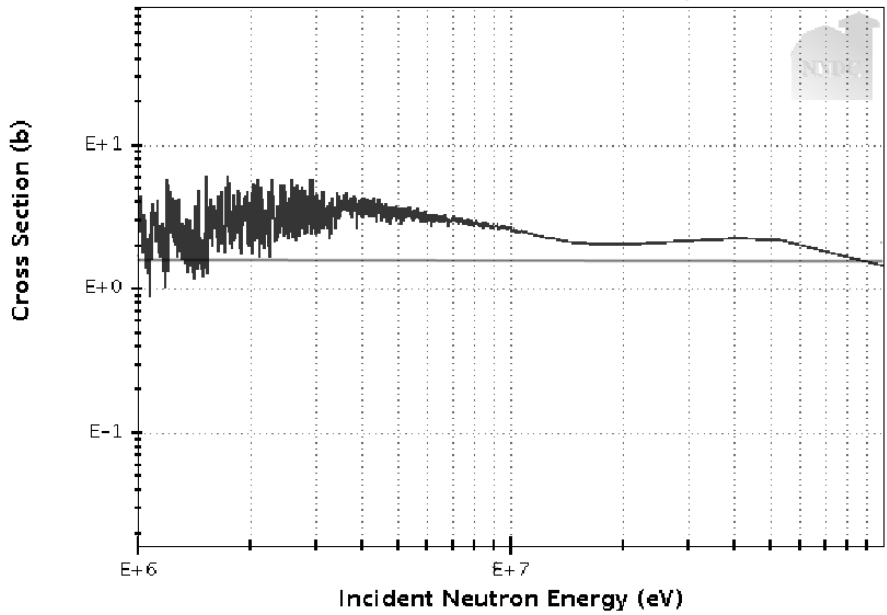}
\caption{ The total \csss for \zn s on $^{40}Ca$ from 1 to 100 MeV,
from the NNDC.
 The
black disc value with $R=A^{1/3}\,1.4 f$ is 1.4 b}
\label{ca40tot}
\end{figure}

\begin{figure}
\includegraphics[width=\linewidth]{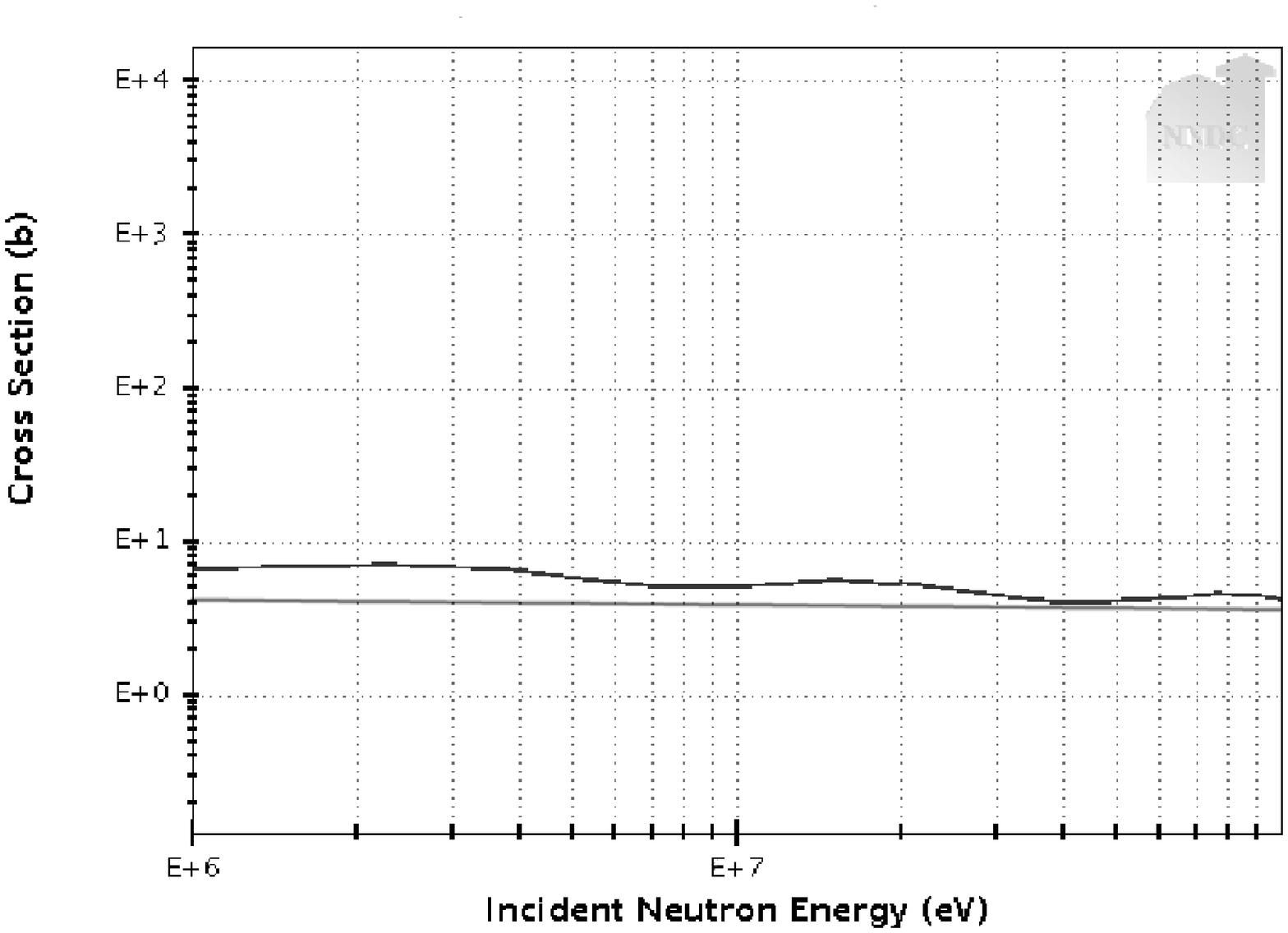}
\caption{ The total \csss for \zn s on $^{184}W$ from 1 to 100 MeV,
form the NNDC.
 The
black disc value with $R=A^{1/3}\,1.4 f$ is 4.0 b}
\label{w184tot}
\end{figure}

 In
Fig\,\ref{o16tot}, Fig\,\ref{ca40tot}, and Fig\,\ref{w184tot}
\cite{boris} we
show as examples the total experimental \csss for the light,
medium, and heavy
weight nuclei oxygen, calcium, and tungsten.
One notes that just above the resonance region the total \csss
becomes smooth  and
tends to be above
the black disc
value, (indicated by the grey line using R as given by \eq{rnuc})
and then slowly descend through it as the \yy increases.

 In the  elastic \sctg
there is a forward
diffraction peak at low momentum transfer $\Delta$. This elastic
\sctg is  concentrated in
a   region characterized  by
\beql{del}
\Delta R\sim 1\, , 
\eeql
where R is the radius of the nucleus. At the same time there is an
inelastic \csss of about the same magnitude.
In the perfect  "black disc" limit the elastic and inelastic \sctg 
cross sections are equal and of magnitude $\pi R^2$. As we shall
discuss below, this
inelastic \sctg could in principle be used as a signal for the
presence of the fast \zns .

\subsection{Validity of the simple `Black disc' model}
The principle features of the simple `Black disc' model result from
the assumption that the nucleus is totally absorbing for \zns \, up
to impact parameter $R$, beyond which there is no \sctg or
absorbtion. This simple assumption has the following consequences:
1) the total \csss is made up of equal elastic and inelastic
 parts, adding up to $\sigma_{tot}=2 \times \pi R^2$, 2) these
features are \yy
independent 3) the form of the elastic \sctg  versus angle or
momentum transfer is given by a simple Bessel function expression,
(see \eq{rexp})
where in particular the form versus momentum transfer is
independent of the \zn \, \yyn. 

As one sees in the examples of Fig\,\ref{o16tot},
Fig\,\ref{ca40tot}, and Fig\,\ref{w184tot} the \yy where the smooth
`optical model' behavior sets in  varies with the target nucleus.
For oxygen the \yy must be above about 10 MeV, for calcium above 4
or 5 MeV, while for tungsten 1 MeV suffices. One also notes from
the plots that with our value for R the total \csss is somewhat
underestimated, particularly at the lower \yys.

 There is a large literature concerning more sophisticated
treatments of the optical model \cite{bd}. There the simple `black'
assumption is replaced by a partial and \yy dependent transparency,
the `hard edge' is softened, modifying the simple radius
assumption, and resonance effects and variation with nucleus may be
included. While all these points are interesting and important they
will not significantly alter our general conclusions, particularly
our main point, as  to be discussed in section \ref{recen} that the
recoil \yys implied by the optical model and characterized by our
parameter $E^o_R$ are in the region anticipated for \dm searches.

\section{Recoil \yys} \label{recen}
It is interesting that if we put \eq{del} in the formula
for the recoil \yy $E_R$

\beql{rec1}
E_R=\frac{\Delta^2}{2M_A}\sim\frac{1}{R^22M_A}\, , 
\eeql
one gets recoils in the region of  interest for many \dm searches.
$M$ is the mass of the
nucleus, $M\approx A\times 1\,GeV$ and for R we use 
\beql{rnuc}
R=A^{1/3}\times1.4\,f.
\eeql

Calling this typical recoil \yy $E_R^o$ 
one obtains ($\hbar, c=1$ units)
\beql{rec}
 E_R^o= \frac{1}{R^22M_A}=\frac{10}{A^{5/3}} MeV\, . 
\eeql
This is the typical \yy characterizing the elastic \sctg of the
``high \yy \zn''. In Table 1 we show the value of $E_R^o$ for some
nuclei.  The numbers run through
the values often considered in  \dm  searches.

The differential \csss is a function of the dimensionless parameter
$x=\Delta R$, and with 
these definitions we can  write 
\beql{delr}
(\Delta R)^2=\frac{E_R}{E_R^o}\, . 
\eeql
The introduction of the dimensionless variable $x$, where
$x^2=\frac{E_R}{E_R^o}$
allows us to describe, within the limitations of the model, all
nuclei in a universal way in terms of their ${E_R^o}$.

 A second interesting point is that in the model this typical
recoil \yy 
for the ``high \yy '' \zns~  depends on the nucleus {\it but not}
on the  \yy of the \zn.  This implies that the whole of the ``high
\yy'' \zn~spectrum contributes in approximately the same way to the
recoils.
This statement is only strictly true insofar as the black disc
model is exactly valid, but it will be seen that the major features
of the model  
hold  approximately  over a wide range of \yyn.

\begin{table}\label{tab}
\begin{center}
\begin{tabular}{|l|l|l|l|}
\hline
$Element$&$A$&R(fermi)& $E_R^o(keV)$\\
\hline
\hline
O&16&3.5 &98\\
\hline
F&19&3.7&74\\
\hline
Na&23&4.0&54\\
\hline
Si&28&4.2&39\\
\hline
Ar&40&4.8&21\\
\hline
Ca&40&4.8 &21\\
\hline
Ge&74&5.9&7.7\\
\hline
I&127&7.0&3.1\\
\hline
Xe&132&7.1 &2.9\\
\hline
W&184&7.9 &1.7\\
\hline
\end{tabular}
\end{center}
\caption{The quantities $R$, $E_R^o$ for various nuclei with mass
number A.}
\end{table}

Finally we note the relation  between \sctg angle and recoil \yy
\beql{erang}
E_R=\frac{\Delta^2}{2M_A}=\frac{(P_{neutron})^2}{M_{A}}(1-cos
\theta)=2 E_{neutron}\frac{m_{neutron}}{M_{A}}(1-cos
\theta)\, ,
\eeql
with $M_A$ the mass of the nucleus.
This simple expression will also be true, replacing ``\zn'' with
``WIMP'', for the elastic WIMP \sctg
when the mass of the WIMP is small compared to that of the nucleus.

 In the `black disc'  limit the \yy spectrum of the recoils is
quite
simple.
The diffraction peak leads to an
$E_R$  spectrum falling off with \yyn, but at least for the lighter
nuclei not so steeply as from
WIMP \sctgn.  The elastic \sctg is
given by a Bessel function expression \cite{serber},
$\frac{d\sigma}{d\Delta^2} \sim \bigl(
\frac{J_1(\Delta R)}{\Delta R}\bigr)^2 $  and  since
$E_R=\Delta^2/2M_A$, one expects
\beql{rexp}
\frac{d\sigma}{dE_R}=2M_A \frac{d\sigma}{d\Delta^2} \sim
\biggl(
\frac{J_1(\Delta R)}{\Delta R}\biggr)^2 \,
,\eeql 
Now since $(\Delta R)^2=E_R/E_R^o$ we  can introduce the convenient
variable $x$, where $x^2=E_R/E_R^o$  and write
\beql{rexpa}
\frac{d\sigma}{dE_R} \sim \frac{d\sigma}{dx^2} \sim \biggl(
\frac{2 J_1(x)}{x}\biggr)^2 ~~~~~~~~~~~~~~~~~~~~x^2=E_R/E_R^o
,\eeql

We include the `2' in the Bessel function expression so it is
normalized to one at
$x=0$.
For small x, $x\lesssim 1$, the expression then  behaves as 
\beql{lox}
\biggl(
\frac{2 J_1(x)}{x}\biggr)^2\ =
1-\tfrac{1}{4}x^2+...=1-\tfrac{1}{4}\frac{E_R}{E_R^o}+...
\eeql

For small $\frac{E_R}{E_R^o}$ this implies a relatively
flat \yy spectrum. Thus for oxygen, between $E_R$ = 10 and 40 keV,
$\frac{d\sigma}{dE_R}$ goes down only by about 8\%. Note again that
it is not necessary to know the \zn\, \yy spectrum to reach this
conclusion, as long as  we have ``high \yy'' \zn s. This is of
course quite different from the rapid fall-off of
WIMP \sctgn, even on light nuclei.
The difference originates in the fact that while for fast \zns\,
the recoil spectrum is determined by the radius of the nucleus, for
WIMPs it is largely determined by the velocity spectrum of the
incoming \plsn, which falls  steeply  above a certain velocity.

 Hence it appears that a characteristic sign of the fast
\zn\, background would be
a {\it rather flat recoil \yy spectrum for the light elements.} 

For $\frac{E_R}{E_R^o}$ larger than one  the black disc elastic
cross section
falls rapidly. For convenience  we show  a few values
in Table 2. One sees that most of the elastic cross section is
contained within $x^2=\frac{E_R}{E_R^o}\lesssim a few$

\begin{figure}
\includegraphics[width=0.7\linewidth,height=\linewidth]
{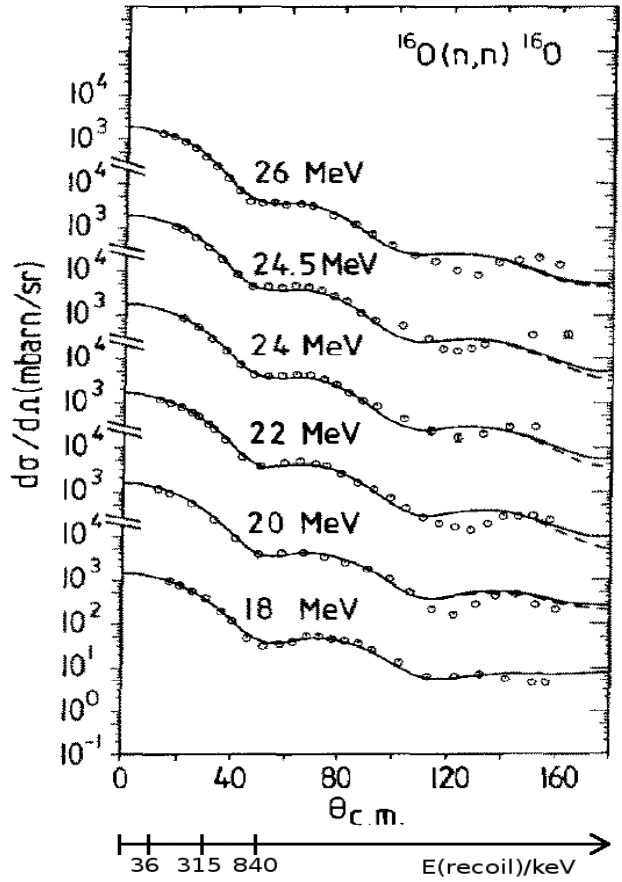}
\caption{ The angular distribution for elastic 
neutron-oxygen \sctg for several \yys from ref~\cite{islam}. The
added
horizontal axis shows, for the
20 MeV case,   $E_R$ for \sctg angles 10, 30 and 50 degrees.  In
the `black disc' limit
 the curves should be the same for all
\yysn, when plotted against $E_R$ instead of angle.}
\label{oxplota}
\end{figure}

\begin{table}\label{tab2}
\begin{center}
\begin{tabular}{|l|l|l|}
\hline
$x$&$E_R/E_R^o$&$\bigl({2 J_1(x)}/{x}
\bigr)^2$\\
\hline
\hline
0&0&1\\
\hline
1.0&1.0&0.77\\
\hline
1.6&2.6&0.51\\
\hline
2.0&4.0&0.33\\
\hline
2.6&7.0&0.13\\
\hline
3.0&9.0&0.051\\
\hline
3.6&13&0.0028\\
\hline
\end{tabular}
\end{center} 
\caption{Some values for the Bessel function expression for the
elastic cross section in the `black disc' model, normalized to 1 at
zero recoil \yyn.}
\end{table}
In Fig\,\ref{oxplota} we show some data on the angular distribution
for oxygen. The optical model shape is quite clear, with a well-
defined low momentum transfer peak and diffractive oscillations,
although at this \yy the effective value for R is slightly smaller
than the one we
are using.
To compare with our variable $\Delta$ we have added an axis showing
the
momentum transfer for 20 MeV \zns. 

\section{Patterns of A Behavior}
Given a certain \zn\, flux in a \dr with  different nuclei,  the
rate will vary in a characteristic way with increasing mass number
A. 
In the black disc limit the total elastic cross section, that is,
integrated over 
all $E_{R}$,  is $\pi R^2$, and so one has

\beql{ela}
\sigma\sim R^2\sim A^{2/3}~~~~~~~~~~~~~~~~all~ E_{R}.
\eeql

However, most \drn s will have a lower \yy threshold  for \dtn and
so will
 miss part  of the diffraction peak. So it is perhaps
also useful to consider the behavior of
$\frac{d\sigma}{dE_{R}}$ at some fixed $E_{R}$, where one has
\beql{elb}
\frac{d\sigma}{dE_{R}}=\frac{\pi}{2} M R^4\biggl(
\frac{2 J_1(x)}{x}\biggr)^2   \sim A^{7/3}\biggl(
\frac{J_1(x)}{x}\biggr)^2 
~~~~~~~~~~~~~~~~~~~x=\biggl(\frac{E_{R}}{E_{R}^o}\biggr
)^{1/2}
\eeql
This formula follows from \eq{rexp} and the use of the optical
theorem point for 
forward \sctg with a purely imaginary amplitude, as one has for the
`black disc':
\beql{opt}
\frac{d\sigma}{d\Delta^2}\bigg\vert_{\Delta=0}=
\frac{1}{16\pi}\sigma_{total}^2\,.
\eeql
In terms of the x variable alone, these relations can also be given
the simple form
\beql{elba}
\frac{d\sigma}{dx^2}=\tfrac{1}{4} \pi R^2\biggl(
\frac{2 J_1(x)}{x}\biggr)^2   
\eeql

 At
fixed $E_{R}$, the  increase of $\sim A^{7/3}$  in \eq{elb} will by
slowed
by the decreasing $E_{R}^o$, which  causes $x$ to move out  in the
Bessel function expression. Since in the `black disc' limit the
recoil spectrum is independent of the incident \zn\, \yy and the
incoming flux is the same for all nuclei in a given experimental
setup, we may take  the rate as approximately proportional to
\eq{elb}. In Table 3 we show, for $E_{R}=20
$  and $30\,keV$,
how \eq{elb} and thus the rate varies with $A$ for some elements.

The interplay of the factors in \eq{elb} leads to a maximum of the
rate at fixed $E_R$ towards the middle of the periodic table. An
interesting  point is that while for lower \yy \zns~ one expects,
on
essentially kinematic grounds, the
\sctg to be predominantly on light nuclei such as oxygen, for
``high \yyn" \zns\,, where the incident \yy is high relative to the
recoil \yyn, this is no longer true.

 All values in the Table are
normalized to that
on oxygen. The entries refer to the rates per nucleus, so for a
given detector account should taken of the relative abundance
of the nucleus in the target material.

For comparison we also show the rates for 10 and  50 GeV WIMPs,
with coherent
$\sim A^2$ interactions, for $E_R=20\, keV.$ For WIMPs not having
the simple coherent interactions, it is not possible to make a
simple statement as to the A behavior since the variation from
nucleus to nucleus will depend on the quantum number constitution
of the WIMP. Indeed it might be possible to unravel these quantum
numbers by studying the A behavior \cite{cooper}.

\begin{table}\label{tab3}
\begin{center}
\begin{tabular}{|l|l|l|l|l|l|}
\hline
$Element$&$A$&{neutron}&neutron&
WIMP&WIMP\\
&        &   {$E_R=$20\,keV}&$E_R=$30\,keV&M=10 GeV& M=50 GeV\\
\hline
\hline
O&16&1&1&1&1\\
\hline
F&19&1.5&1.5&1.3&1.8\\
\hline
Na&23&2.2&2.2&1.6&3.3\\
\hline
Si&28&3.4&3.3&1.8&6.7\\
\hline
Ar&40&7.0&6.4&1.1&19\\
\hline
Ca&40&7.0 &6.4&1.1&19\\
\hline
Ge&74&19&13&$\sim\,$0&93\\
\hline
I&127&20&5.1&$\sim\,$0&200\\
\hline
Xe&132&18 &3.9&$\sim\,$0&240\\
\hline
W&184&2.6 &1.6&$\sim\,$0&230\\
\hline
\end{tabular}
\end{center}
\caption{ Variation of the differential \sctg rate per unit \yy 
  over  various nuclei in the `black disc' limit, \eq{elb}, 
at $E_{R}=$20\,keV and  $E_{R}=$30\,keV. 
 For comparison the same
rate for a   coherently \sctg WIMP at $E_R=20\,keV$  for masses 
 10 and 50 GeV is also shown. One notes  different patterns of A
behavior for \zns\, and WIMPs.  All
values are per nucleus and normalized to that for oxygen.} 
\end{table}

\eq{elba} may be used to estimate the event rate in an acceptance
window for a given setup and an assumed \zn\, flux. 
Bessel function relations can be used to estimate what fraction of
the elastic
\csss $\pi R^2$ is in a certain interval.  Thus, for
example,  assume a $\rm CaWO_4$ \dr accepting events  on the
oxygen between 10\,keV and 40\,keV. From Table 1 this correspopnds
to
$x^2={E_R/E_r^o}$  between 0.11 and 0.42. On the other hand the
total
elastic \csss according to \eq{elba} is proportional to
$\int_0^\infty (2 J_1(x)/x)^2 dx^2$. Now there is the
``orthonormality relation''  \cite{wiki} $\int_0^\infty
(2J_1(x)/x)^2 dx^2=4$, where we have normalized as in \eq{lox}.
According to \eq{lox} one may take $(2J_1(x)/x)^2$ as
approximately equal to one in this interval. Thus we have a
fraction of the elastic \csss $0.31/4=0.078$ in the interval
10\,keV to 40\,keV.
Since $\pi R^2=0.39\,b$, there is thus a \csss of 0.030 b in the
interval.
 
For  a flux of say $1\times
10^{-7}/cm^2 s$ for the ``high \yy'' \zns, this leads to a rate of
$2\times 10^{-3}$/kg\,day for $\rm CaWO_4$. Some information on
underground \zn\, fluxes may be  found in Reference \cite{nund}.
Naturally given the \zn\, flux at a given location, the \zns\, must
be further be propagated through the shielding and materials of the
setup in question, and there is  the further question of \zns\,
originating  in these materials.

\section{Inelastic Processes}

An important difference between \zns \, and WIMPs is of course that
\zns\, are
strongly interacting and WIMPs weakly interacting. This implies
that,
depending on the material and geometry of the \drn, multiple \sctg
may serve as a signal for  \zn s.

A further  significant difference between the MeV \zns~we discuss
here and
the assumed galactic WIMPs is the low \yy of the WIMP. 
The WIMP is expected to have a small velocity,  typical of objects
in
the Milky Way, $v\sim 2\times
10^{-3}$ (c=1 units).
Then even, say, a mass 50 GeV WIMP only has an \yy
 $\hf mv^2 \sim 100\, keV$, and a `light WIMP' with M=10 GeV only
$\sim 20\, keV$. Such \yys are generally not enough to induce
significant inelastic processes on most nuclei.

\begin{figure}[h]
\includegraphics[width=\linewidth]{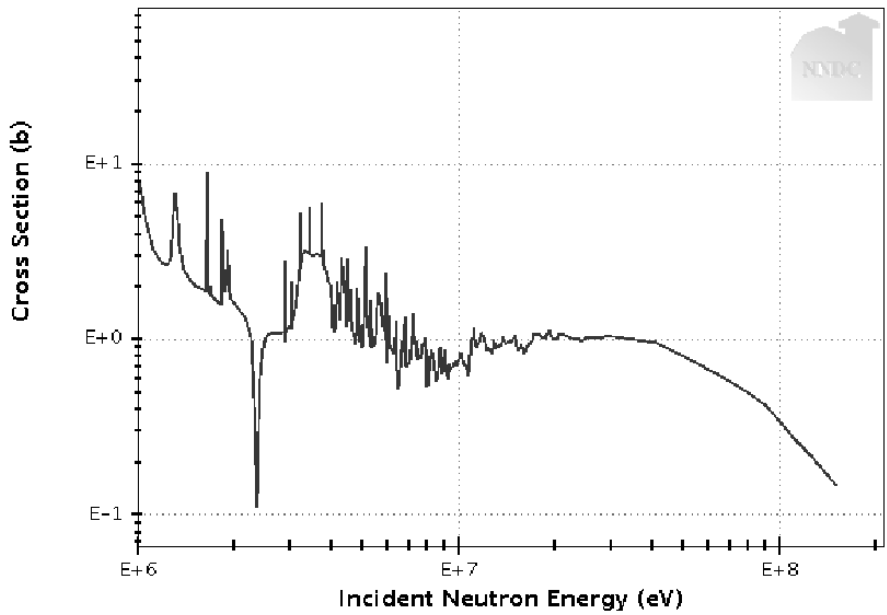}
\caption{ The total \csss for elastic \sctg on  $^{16}O$, from the
NNDC. As compared to the total \csss in  Fig\,\ref{o16tot}, this
elastic \csss includes only processes where the nucleus undergoes
no breakup or excitation. One notes that the simple optical model
relation $\sigma_{elastic}\approx \hf
\sigma_{total}$ is approximately correct. }
\label{o16el}
\end{figure}

On the other hand,  \zns\, with MeVs and above typically have many
inelastic reations and
the inelastic reactions are expected at close to
the same rate as the elastic \sctgn. In the `black disc' limit' one
has
$\sigma_{elastic}\approx \sigma_{inelastic}\approx
\hf \sigma_{total}$. Fig\,\ref{o16el} \cite{boris}  for the elastic
\sctg on
oxygen, when
compared with Fig\,\ref{o16tot}, shows this is approximately true.

In experiments with two-signal readout designed to distinguish
electromagnetic backgrounds from nuclear events, such as CRESST,
Edelweiss, CDMS, Xenon.. . these inelastic events will typically
appear as a high \yy signal in the nuclear event band.
Depending on the nuclei in question,
the most common inelastic  channels  will be nuclear excitation,
leading to a $\gamma$ ray, or  nuclear breakup.  In
breakup
common  nuclear fragments  are further \zns\, and  $\alpha$ \plsn,
while for  nuclear excitation followed by $\gamma$ rays
electromagnetic \yy is present. Therefore inelastic events will
generally have a relatively high  \yy deposit, compared to the
elastic recoils either from WIMPs or \zns.  Depending on
the nuclei present in the \drn, these inelastic events could also
appear  on nuclei   other
than those exhibiting the small  recoils. Thus associated
with a fast \zn~ flux there should be a characteristic set of
events
with quite different properties from the small elastic \sctg
recoils.
But these should be absent for WIMPs.

\section{ Conclusions}
We have pointed out that the difficult-to-shield fast \zn s will
give elastic  nuclear recoils of the same type as sought for in \dm
searches. We explain, however, that it should be possible 
 to use evidence
internal to the \dr in establishing the presence of a \zn\,
background and so 
not to  be only dependent upon standard assumptions and  Monte
Carlo
estimates.  We have suggested some methods of doing this,
 including the pattern of recoil \yys, the rates on different
nuclei through the periodic table (see section 4) and the presence
of inelastic nuclear processes at about the same level as elastic
nuclear recoils (see section 5). 
 In
this respect  it is thus desirable  to have \dr systems able to
distinguish more than one type of recoiling nucleus \cite{cresst}.

 For
light to medium weight nuclei the \zn\, induced events should show
a rather flat
spectrum with respect to  recoil \yy $E_R$. 
At a fixed recoil \yyn, the variation of the rate with respect to
changing the nucleus is
 different for \zns\, and WIMPs as explained in
connection with Table\,3.

%\ref{tab3}.

Neutrons can of course have multiple interactions, and fast \zns\,
as opposed to WIMPs can induce inelastic reactions, and at a
relatively high rate.

 The great difficulty,
 as with all aspects of \dm searches, is that at present only a
handful of events is available and that a
detailed investigation of these points  requires  extensive
data,  not easy to come by in very low rate experiments.

\section{Acknowlegements}
I would like to thank F. Proebst and J. Schmaler  for many
discussions on this subject. The WIMP rates in Table 3 were
calculated using the  program  `fancylimit' developed R. Lang and
J. Schmaler.

I would also like to thank Boris Pritychenko for his  help in
understanding the use of the \zn\, data compilations of the NNDC.

\newpage

\end{document}